 \documentclass[journal, comsoc]{IEEEtran}
\usepackage[a4paper,bindingoffset=0.0in,
left=.65in,right=.6in,top=.9in,bottom=.9in]{geometry}
\usepackage{gensymb}
\usepackage{cite}
\usepackage{amsmath,amssymb,amsfonts}
\usepackage{graphicx}
\usepackage{xcolor}
\usepackage{kotex} 
\usepackage{textcomp, gensymb}
\usepackage{stfloats}
\usepackage{float}
\usepackage{wrapfig}
\usepackage{xcolor}
\usepackage{colortbl} 
\usepackage{color,soul}
\usepackage{ragged2e}
\usepackage{url}
\usepackage{verbatim}
\usepackage{graphicx}
\usepackage{cite}

\usepackage{multirow}
\usepackage{array,boldline, makecell,booktabs}
\usepackage{longtable} 

    \usepackage[group-separator={,},group-minimum-digits=4]{siunitx}

\begin{document}
%

\title{A Quick Guide to Quantum Communication}
%

\author{Rohit Singh, \textit{Member, IEEE}, Roshan M. Bodile, \textit{Member, IEEE} \thanks{Authors are with the Department of Electronics and Communication, Dr B R Ambedkar National Institute of Technology Jalandhar, India (e-mail: \{rohits, mukindraobr\}@nitj.ac.in) }}

%
%

\markboth{}%
{Shell \MakeLowercase{\textit{et al.}}: Bare Demo of IEEEtran.cls for IEEE Journals}

\maketitle

\begin{abstract}
This article provides a quick overview of quantum communication, bringing together several innovative aspects of quantum enabled transmission. We first take a neutral look at the role of quantum communication, presenting its importance for the forthcoming wireless. Then, we summarise the principles and basic mechanisms involved in quantum communication, including quantum entanglement, quantum superposition, and quantum teleportation.  Further, we highlight its groundbreaking features, opportunities, challenges and  future prospects.
\end{abstract}


\IEEEpeerreviewmaketitle

\vspace{-3mm}

\section{Introduction}
\label{sec:1}
Over the past few decades, wireless communication
has witnessed tremendous advancements \cite{irs}.  Yet, the forthcoming wireless network demands even faster, more efficient, and secure information exchange, where quantum communication appears to be a promising solution. The quantum world, with its counter-intuitive properties, has gained significant attention from physicists for decades. Quantum mechanics, the fundamental theory governing the behaviour of particles, creates a pathway to a wide range of technological advancements. Indeed, the most promising implementation of quantum mechanics lies in its potential to revolutionize the way we communicate with each other. Quantum communication precisely makes use of two fundamental theories namely \textit{quantum entanglement} and \textit{quantum  superposition}. 

\begin{figure}
    \centering
    \includegraphics[width=1\linewidth]{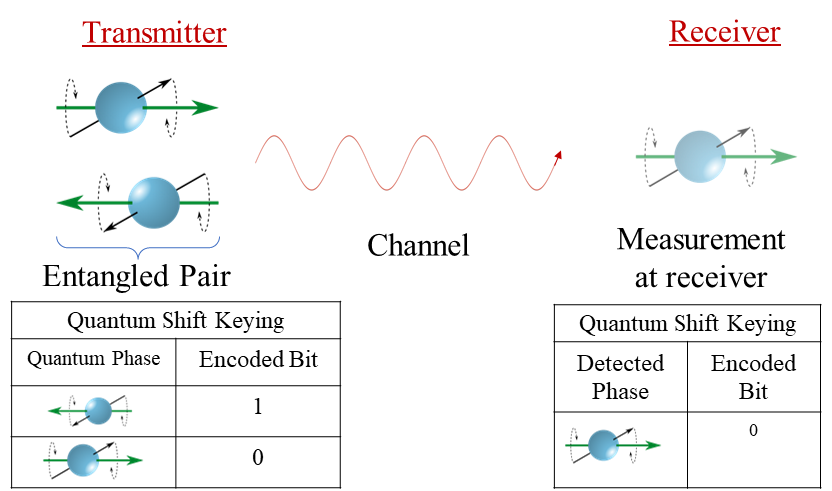}
    \caption{An illustration of quantum communication, depicting encoding and decoding.}
    \label{fig1}
\end{figure}

Indeed, the concept of quantum entanglement was initially postulated by the renowned physicist \textit{Albert Einstein}, popularly known as ``spooky action at a distance''. Specifically, entanglement is a phenomenon which interconnects qubits in such a way that the state of one is linked to the state of another, irrespective of the travelled distance \cite{book}. Whereas, superposition is another phenomenon, in which quantum can exist in multiple states simultaneously. This duality conservation of spinning quantum pairs is mainly utilized for information encoding and is known as Quantum Shift Keying (QSK). Unlike classical communication which uses classical bits, quantum communication uses quantum bits, popularly known as \textit{qubits}. Fig. \ref{fig1}  illustration of information transmission through qubits.  Apart from new definitions of information exchange, quantum-enabled transmission provides several other complementary features,  with a particular research focus on enhancing network security.  The phenomenon of entanglement is utilized for producing encryption keys, known as Quantum Key Distribution (QKD). QKD (defined in section \ref{qkd}) is a secure communication protocol that ensures the confidentiality of cryptographic keys by leveraging the principles of quantum mechanics.

Reviewing the principle, benefits and potential applications, this article provides a quick overview of quantum communication, including \textit{a)} the role of quantum communication towards the evolution of communication technologies beyond 6G, \textit{b)} presenting the fundamental principles of quantum communication, under the realm of quantum entanglement, quantum superposition, and quantum teleportation \textit{c)} further, highlighting the groundbreaking features and opportunities through quantum communication, including \textit{Quantum Shift Keying}  in the context of 6G and beyond, \textit{d)} besides, this article presents current state and challenges of quantum communication in wireless technology, \textit{e)} finally, the future prospects have been given to promote research in this domain.

\section{Principles and Significance}
Quantum Communication, being a field of applied physics, explores the principles of quantum mechanics to transmit and process information. This section summarises key technology pillars of quantum communication.


\subsection{Quantum Entanglement}
\begin{figure}
    \centering
    \includegraphics[width=0.85\linewidth]{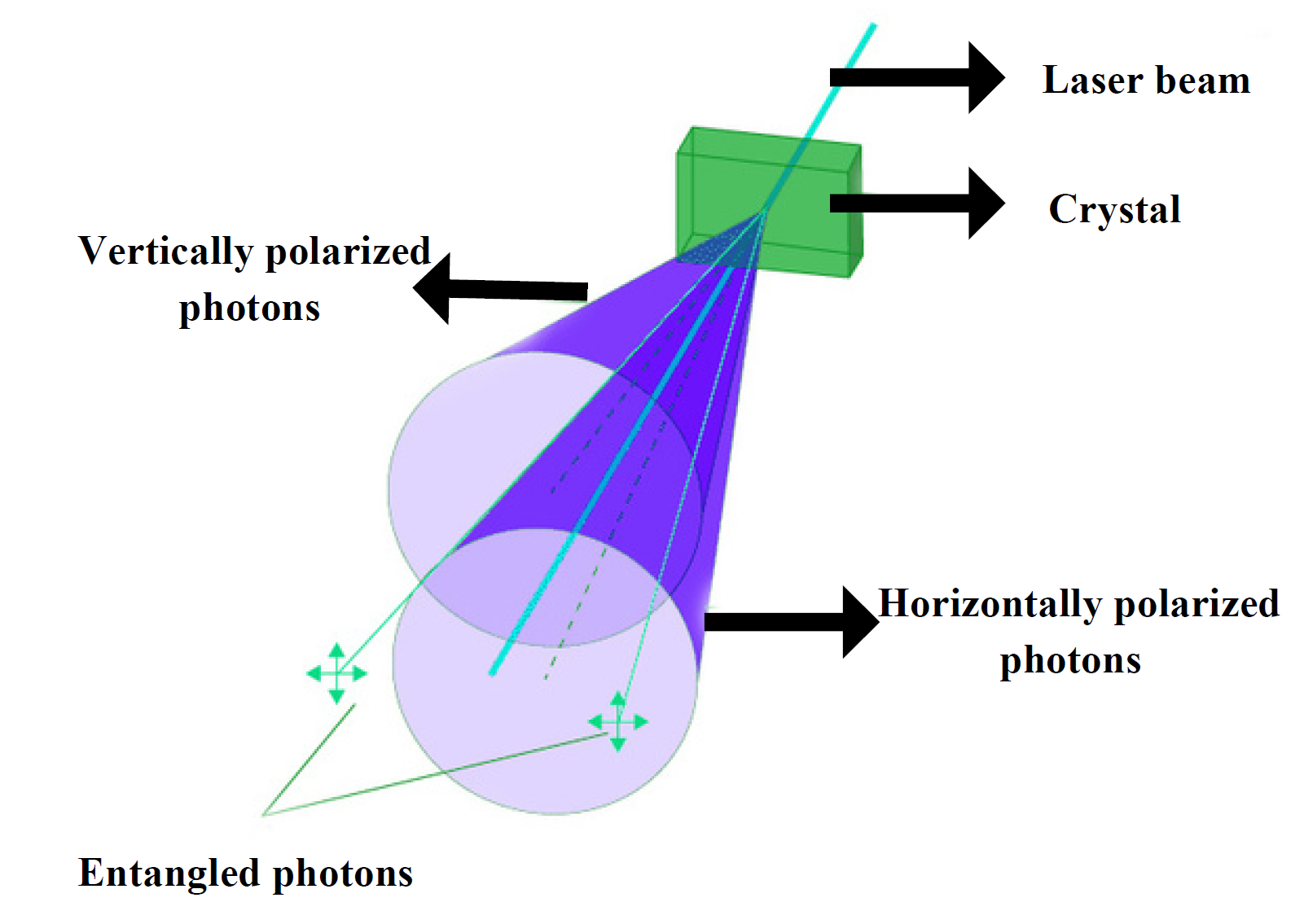}
    \caption{An illustration of entangled particles and impact of measurement on their states.}
    \label{figtang}
\end{figure}

Quantum entanglement is the core principle\footnote{This intrinsic connection is mainly utilized for information cloning, leading to the development of quantum shift keying.} stated as ``\textit{when two particles interact, they can become entangled such that their quantum states are linked, even when separated by large distances}''. As depicted in Fig. \ref{figtang}, measurements on one entangled particle instantaneously affect the state of the other, enabling correlations between measurements that cannot be explained by classical physics \cite{ent}. Quantum entanglement appears due to the principle of particle correlation, in which the state of one particle is linked to the state of another, regardless of the distance between them.

Not only the pairs of particles, entanglement may also involve multiple particles, known as multi-partite entanglement. Nevertheless, it brings both opportunities and challenges. Though the inclusion of multiple particles opens possibilities for sophisticated quantum communication protocols and quantum computation algorithms, the complexity of entangled states increases with particles involved. In quantum physics, the states of entangled particles are described by a joint wave-function, and changes in the state of one particle instantaneously affect the state of its entangled counterpart. Owing to distinct features of quantum entanglement, researchers are exploring its role in various applications including quantum computing, where qubits can exist in superposition states, enabling the parallel processing of vast amounts of information. Overall, quantum entanglement is not just a theoretical idea and leads to the creation of innovative technologies like quantum cryptography, quantum teleportation, and quantum computation.

\subsection{Quantum Superposition}
\begin{figure}
    \centering
    \includegraphics[width=0.85\linewidth]{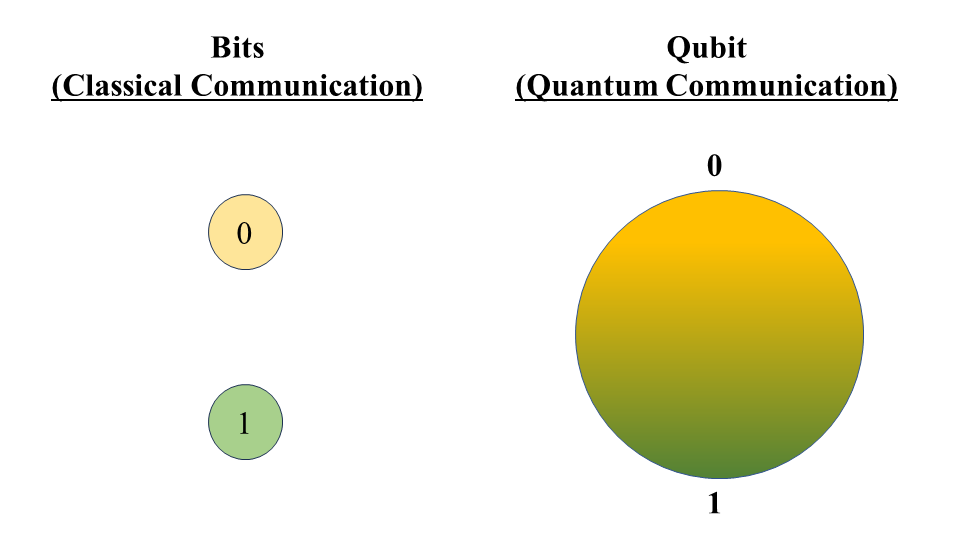}
    \caption{An illustration of classical information mapping and superposition of information in qubits.}
    \label{figsup}
\end{figure}

Quantum superposition is another key principle in quantum mechanics that challenges classical intuitions, allowing particles to exist in multiple states simultaneously, as depicted in Fig. \ref{figsup}. Moreover, the experimental validation of quantum superposition is often justified through double-slit experiments. When particles are directed through two slits, they create an interference pattern on a detection screen, revealing that the particle travels through both slits simultaneously. This interference phenomenon is a direct consequence of the superposition of different quantum states. Thus, unlike classical physics systems with definite states, superposition enables particles to be in multiple states at once, represented by a linear combination of their individual states. This unique feature arises from the wave nature of quantum entities, described by wave-functions. 


Similar to entanglement, particles can be superpositioned in a combination of many states, a property exploited in quantum computing. Also, beyond quantum computing, it plays a major role in quantum cryptography. This is enabled through the manipulation of quantum states, offering the secure exchange of information via QKD protocols.  Until a measurement is made , the particle is in a superposition of these states, highlighting the probabilistic nature of quantum systems.

\subsection{Quantum Teleportation}
Quantum teleportation allows the transfer of quantum information from one location to another without the physical movement of particles. As depicted in Fig. \ref{fig1}, quantum teleportation relies on the principles of quantum entanglement via the following process: \textit{a)} the fundamental steps of quantum teleportation involve the preparation of an entangled pair, known as an Einstein-Podolsky-Rosen (EPR) pair, by sharing them between two distant locations, \textit{b)} one of the entangled particles is then entangled with the quantum information to be transmitted\footnote{The state of the original particle is destroyed, and the information is transmitted to the distant entangled particle.}, \textit{c)} at the receiver end, the party can then use the transmitted information to recreate the original quantum state, completing the teleportation process.\footnote{Despite the name ``teleportation'', this process only transfers the information associated with the quantum state, not the physical particle itself.} 

Similar to the how classical shift keying encodes the information either in phase, amplitude, or frequency, QSK applies quantum twist by manipulating quantum states on the basis of encoded information. Unlike classical phase shift keying (PSK) where variations in the phase of carrier waves serve as the basis for encoding bits, QSK uses quantum states, such as the polarization states of photons, which serves a carrier of information. Besides, quantum enabled transmission provides complementary benefits in terms of data security through QKD, preventing the unauthorized replication of information.

\textit{Einstein-Podolsky-Rosen (EPR): } The EPR paradox is rooted in the entanglement phenomenon. It is based on the fact that the correlated measurements on entangled particles can be made with certainty, even without disturbing the particles themselves. This led to the formulation of EPR, which states that if the position and momentum of one particle are known, the position and momentum of the entangled particle can be predictable.

\section{Ground Breaking Features, Challenges, and Future Prospects}
Previous section explored how the transmission through quantum enabled framework redefines the ways we communicate, though their are also a few associated challenges \cite{chal}. This section covers ground breaking features, challenges, and future prospects, associated to quantum communication. 
\begin{figure}
    \centering
    \includegraphics[width=0.9\linewidth]{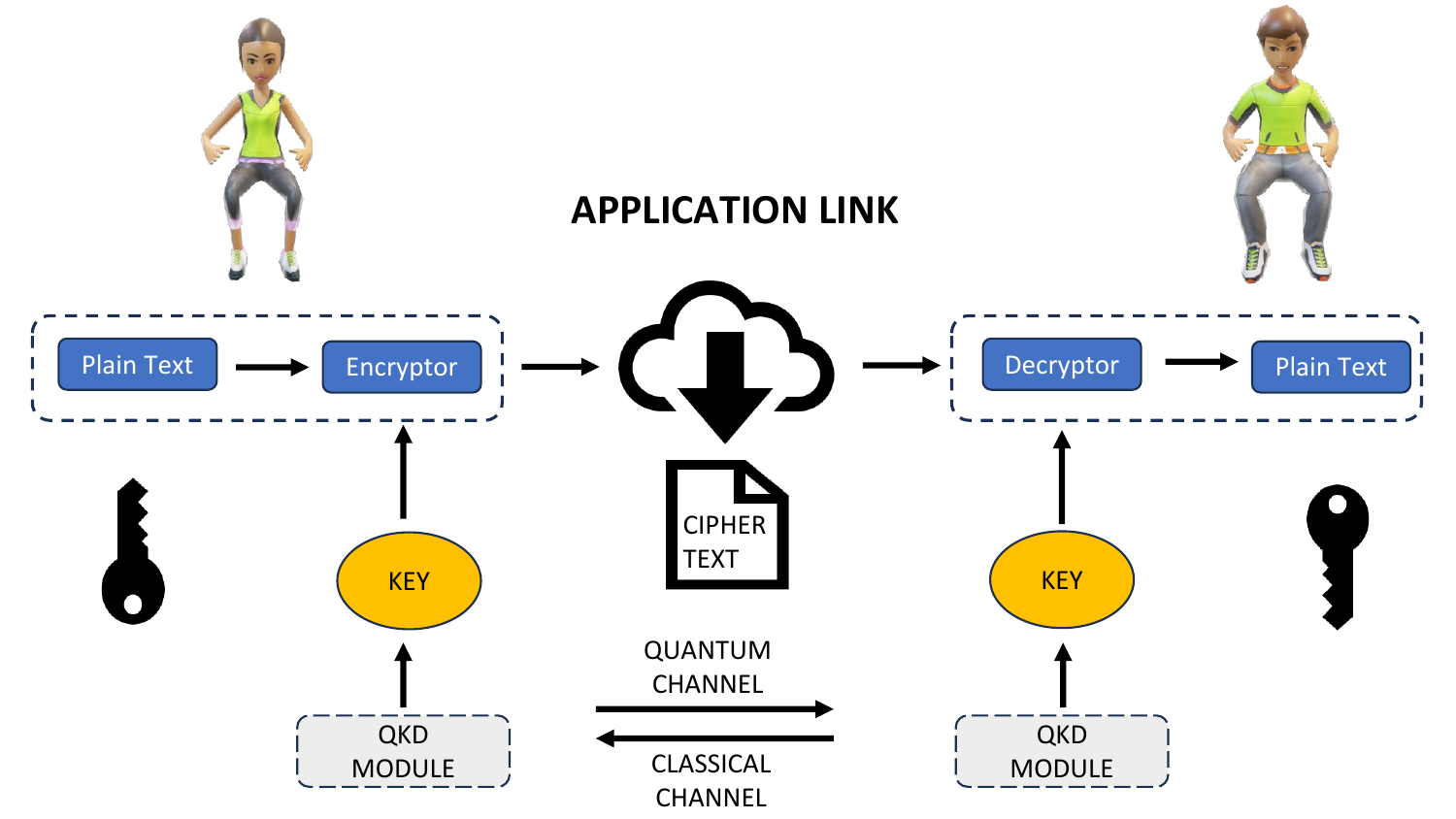}
    \caption{An illustration of data encryption through QKD.}
    \label{fig4}
\end{figure}

\subsection{Quantum Key Distribution: Redefining Data Encryption}\label{qkd}

By utilizing the phenomenon of quantum twist and correlation, QKD introduces a fundamental property of quantum mechanics known as the \textit{no-cloning theorem}. This theorem states that it is nearly impossible to create an identical copy of an unknown quantum state. Consequently, any unauthorized attempt to intercept the transmitted keys in QKD will unavoidably disturb the entangled particles, thereby alerting legitimate users to the presence of eavesdroppers. This inherent feature of QKD ensures that any illicit acquisition of the key by eavesdroppers cannot occur without being detected.
The unbreakable link mentioned in the user’s text serves as the fundamental basis for quantum cryptography, which in turn establishes the groundwork for ensuring secure quantum communication.

\subsection{Quantum Communication and 6G and Beyond}
Based on the principles of quantum mechanics, quantum communication holds immense potential for the upcoming wireless generations. This section covers potential advances through the inclusion of quantum theory under the realm of \textit{6G}. Researchers are making significant efforts to prove the realization of quantum-based transmission, though yet there is no evidence confirming the consensus on the use of quantum communication for the upcoming wireless generation. The joint utilization of quantum-assisted transmission intends to provide numerous benefits in 6G technologies including, e.g., quantum-enhanced communication channels. Both academia and industry have well studied the theoretical aspects of quantum teleportation, however, the translation of these concepts into practical implementation is being explored.

\subsection{Challenges and Future Prospects}
Apart from the numerous benefits, quantum communication comes up with several associated challenges and Futuristic prospects. Some of them are discussed below as:  

\subsubsection{Integration with Classical Communication Networks}
It is nearly impossible to replace the existing communication system as the quantum communication relies on completely different setup. Thereby, the integration with classical network encompasses both the challenges and future opportunities. Future advancements include the possibilities of seamless integration with classical communication, leveraging the benefits of both the worlds.

\subsubsection{Quantum Repeaters for Long-Distance Communication}
Scaling up the transmission across boundaries is a significant challenge, though the successful teleportation of quantum states has been achieved across a few kilometers using optical fibres. As the information lies in orientation, preserving entanglement across substantial distances is quite difficult. Specifically, entanglement is susceptible to losses with the transmission length through optic channels that limits the effective range of quantum communication \cite{lim}. Nevertheless, these challenges may be addressed through the successful recreation of entanglement links over shorter segments.\footnote{This intends to work as chain by further connecting these links to extend quantum communication capabilities over longer distances.}

\subsubsection{Quantum Satellite Communication}
Quantum satellite communication leverages entangled photon pairs generated on board the satellite and transmitted to ground stations \cite{sat}. The use of benefits in terms of global coverage, while the absence of fiber optic mitigates the issues of signal decoherence. Besides, satellites can also be use as the third party sharing of secret keys among ground stations. Moreover, the satellite integration bring several complementary benefits in terms of distributed quantum computing, paving the way for advanced applications in quantum information processing on a global scale.

\section{Conclusion}
This article provides a quick overview of quantum communication including basic principles, benfits, challenges and future direction. Overall, quantum communication holds great promise and intends to offer a more efficient and stable transmission of information in the form of quantum bits, commonly referred to as qubits. Besides, quantum communication is intended to evolve in the near future a lot via optimizations, machine learning and cryptography. Moreover, as the technology matures and becomes more accessible, it is expected to be increasingly used in industries such as finance and healthcare, where it can be used to analyse large amounts of data and make more accurate predictions.

\bibliographystyle{IEEEtran}


\end{document}